\documentclass[%
 reprint,
 amsmath,amssymb,
 aps,
]{revtex4-2}
\usepackage{graphicx}
\usepackage{dcolumn}
\usepackage{bm}
\usepackage{hyperref}
\usepackage{amsmath}
\hypersetup{hypertex=true,
            colorlinks=true,
            linkcolor=blue,
            anchorcolor=blue,
            citecolor=blue}

\begin{document}

\preprint{APS/123-QED}

\title{Nuclear mass predictions based on convolutional neural network}

\author{Yanhua Lu, Tianshuai Shang, Pengxiang Du, Jian Li}
\email{E-mail:jianli@jlu.edu.cn}
 \affiliation{College of Physics, Jilin University, Changchun 130012, China.}
 \author{Haozhao Liang}
\affiliation{Department of Physics, Graduate School of Science,	The University of Tokyo, Tokyo 113-0033, Japan}
\affiliation{RIKEN Interdisciplinary Theoretical and Mathematical Sciences Program, Wako 351-0198, Japan}
\author{Zhongming Niu}
\affiliation{School of Physics and Optoelectronic Engineering, Anhui University, Hefei 230601, China}

\date{\today}

\begin{abstract}
A convolutional neural network (CNN) is employed to investigate nuclear mass. By introducing the masses of neighboring nuclei and the paring effects at the input layer of the network, local features of the target nucleus are extracted to predict its mass. Then, through learning the differences between the experimental nuclear masses and the predicted nuclear masses by the WS4 model, a new global-local model (CNN-WS4) is developed, which incorporates both the global nuclear mass model and local features. Due to the incorporation of local features, the CNN-WS4 model achieves high accuracy on the training set. When extrapolating for newly emerged nuclei, the CNN-WS4 also exhibits appreciable stability, thereby demonstrating its robustness.
\end{abstract}

\maketitle


\section{\label{sec:1}introduction}

Nuclear mass is one of the basic quantities in nuclear physics, providing rich structural information, such as shell effect~\cite{ramirez2012direct,wienholtz2013masses} and deformation~\cite{hager2006first,de2017nuclear}. It is widely used to extract nuclear effective interactions, such as mean-field effective interactions~\cite{brown1998new,lalazissis2005new,zhao2010new} and pairing interactions~\cite{satula1998odd}. Additionally, it is also a crucial input in nuclear physics for understanding the energy production in stars and the origin of elements in the universe by determining the reaction energies
of all involved nuclear reactions~\cite{burbidge1957synthesis,mumpower2016impact,bethe1939energy}. Therefore, nuclear mass plays a vital role in understanding nuclear structure and studying the astrophysical nucleosynthesis.

Due to the importance of nuclear mass, there has been a persistent pursuit to obtain more precise nuclear mass. Many advancements have been achieved in the experimental measurements of nuclear mass~\cite{franzke2008mass,sun2015toward}. AME2020 provides a comprehensive overview of nuclear mass measurements, evaluating and recommending the masses of 3557 nuclei~\cite{huang2021ame,wang2021ame}. On the theoretical side, different kinds of nuclear mass models have been developed~\cite{lunney2003recent}. In 1935, Weizs$\ddot{\rm{a}}$cker~\cite{weizsacker1935theorie,bethe1936nuclear} proposed for the first time a semi-empirical formula for the masses of nuclei, namely the famous Bethe-Weizs$\ddot{a}$cker (BW) mass formula, which predicted the mass with an accuracy of about 3 MeV. The semi-empirical formula belongs to the macroscopic model and can describe the mass of most nuclei well. By incorporating the microscopic shell correction term into this model, macroscopic-microscopic models, such as the finite range liquid droplet model (FRDM)~\cite{moller2012new} and the Weizs$\ddot{\rm{a}}$cker-Skyrme (WS) model~\cite{wang2014surface}, have been developed. The corresponding prediction accuracy has been significantly improved to approximately 0.3 MeV. The microscopic mass models based on the nuclear density functional theory have also been developed, e.g., the
relativistic mean-field model~\cite{RCHBMENG2006470,VRETENAR2005101,mengjiePhysRevC.73.037303,lianghaozhaoPhysRevLett.101.122502,niuzhongmingPhysRevC.87.051303,niuzhongmingPhysRevC.95.044301,XIAxuewei20181,relativisticmeanfieldmengjie2005,hua2012comparative,pena2016relativistic} and the nonrelativistic
Hartree-Fock-Bogliubov (HFB) model with Skyrme~\cite{goriely2016further} or
Gogny~\cite{goriely2009first} force. They are generally believed to provide better reliability in extrapolation, although their accuracy in predicting experimentally known masses is currently lower than those of macroscopic-microscopic models~\cite{Zhao2012PRC}.

The theoretical models mentioned above are commonly referred to as global theoretical models. In contrast, the models based on the assumption of local interaction and connecting the physical quantities of adjacent nuclei are called local type theoretical models~\cite{jiang2012predictions,garvey1966new,jiang2010nuclear,fu2011description}. The local mass models have unique advantages in predicting nuclear mass, especially in regions where experimental nuclear mass data are known. The local mass models primarily include the Garvey-Kelson local mass relation (GK relation)~\cite{garvey1966new}, the mass relation based on neutron-proton interaction ($\delta V_{in-jp}$ relation)~\cite{jiang2010nuclear,fu2011description}, and the Audi-Wasptra extrapolation method developed during the compilation of the AME data table~\cite{huang2021ame,wang2021ame}. In the region where the experimental data are known, the local type models can achieve high accuracy. For instance, local mass relations achieve an accuracy of approximately 0.2 MeV~\cite{jiang2012predictions}.

Despite significant improvements in the prediction accuracy of nuclear mass models, theoretical models still fail to meet the research demands for exotic nuclear structure and astrophysical nucleosynthesis~\cite{tiandachuanmass2022}. For example, the rapid neutron-capture process ($r$-process) is a prominent area in physics, demanding a nuclear mass prediction accuracy of about 100 keV~\cite{mumpower2016impact}. Hence, there is a considerable scope for improving the existing nuclear mass models. Machine learning (ML) is one of the most popular approaches to complex problems due to its powerful learning capabilities. In the field of nuclear physics, machine learning plays a crucial role with significant applications~\cite{RevModPhys.94.031003,zongshuhewanbing2023}, including the prediction of nuclear ground-state or excited-state properties, such as charge radius~\cite{utam2016nuclear,ma2020predictions,wu2020calculation,akkoyun2013artificial}, excited states~\cite{akkoyun2022estimations,lasseri2020taming,WANG2022137154,BAI2021136147}, $\alpha$ decay~\cite{rodriguez2019alpha,rodriguez2019bayesian}, $\beta$ decay~\cite{niu2019predictions,Li_2024}, charge density~\cite{shang2022prediction,PhysRevC.108.034315,shang2024global}, density functional~\cite{YANG2023137870}, nuclear level density~\cite{PhysRevC.109.044325}, and ground-state magnetic moments~\cite{yuan2021magnetic}. Especially, machine learning has been widely applied to nuclear mass predictions~\cite{neufcourt2020quantified,utama2016nuclear,utama2017refining,niu2018nuclear,neufcourt2018bayesian,neufcourt2019neutron,neufcourt2020beyond,niu2019comparative,sprouse2020propagation,pastore2020impact,liu2021improved,lovell2022nuclear,mumpower2022physically,WUxinghuikrr128kev,niu2022nuclear,du2023examination,PhysRevC.109.034318}. 

Although there are many studies using machine learning to predict nuclear masses, most of them achieve an accuracy of only around 200 keV. To overcome this bottleneck, it is necessary to consider more physics, as demonstrated by some successful studies~\cite{WUxinghuikrr128kev,niu2022nuclear}. For example, in Ref.~\cite{WUxinghuikrr128kev}, employing an extended kernel ridge regression approach to predict the nuclear mass, the root-mean-square mass deviation can be reduced to 128 keV. In Ref.~\cite{niu2022nuclear}, by keeping the known physics in various sophisticated mass models and performing the delicate design of neural networks, the Bayesian Machine Learning (BML) model is developed and an accuracy of 84 keV is achieved in the known region. However, the physics considered in this paper is a novel approach, which considers both the global nuclear mass model and local features. To this end, the convolutional neural network (CNN) is one of the methods to implement this approach. The final results also confirm this idea and achieve a new level of accuracy of 70 keV.

 CNN offers significant advantages in image processing due to its capability to efficiently capture and process the spatial structure of the images. It has been employed to address challenges related to density functional theory~\cite{mills2017deep,ryczko2019deep}. In the field of nuclear physics, CNN has recently been used to acquire knowledge about nuclear charge radii~\cite{su2023progress,cao2023predictions}. 
 
 This paper represents the first use of convolutional neural networks for predicting nuclear mass. We start by directly predicting nuclear mass using CNN. Then, we consider the existing theoretical model to predict the differences between experimental values and the theoretical mass predictions. After considering the WS4 global nuclear mass model~\cite{wang2014surface} based on the convolutional neural network, we construct a global-local model, which considers both the global nuclear mass model and local features. The predicting accuracy has been significantly enhanced, where CNN demonstrates a strong ability to extract local features of target nuclei.

The paper is organized as follows. The details of CNN and data processing are discussed in Sec.~\ref{sec:2}. Section~\ref{sec:3} provides the results and discusses the learning and extrapolation abilities of the proposed models. A summary and perspectives are given in Sec.~\ref{sec:4}.

\section{\label{sec:2}Theoretical framework}

\begin{figure*}
\centering
\includegraphics[width=16cm]{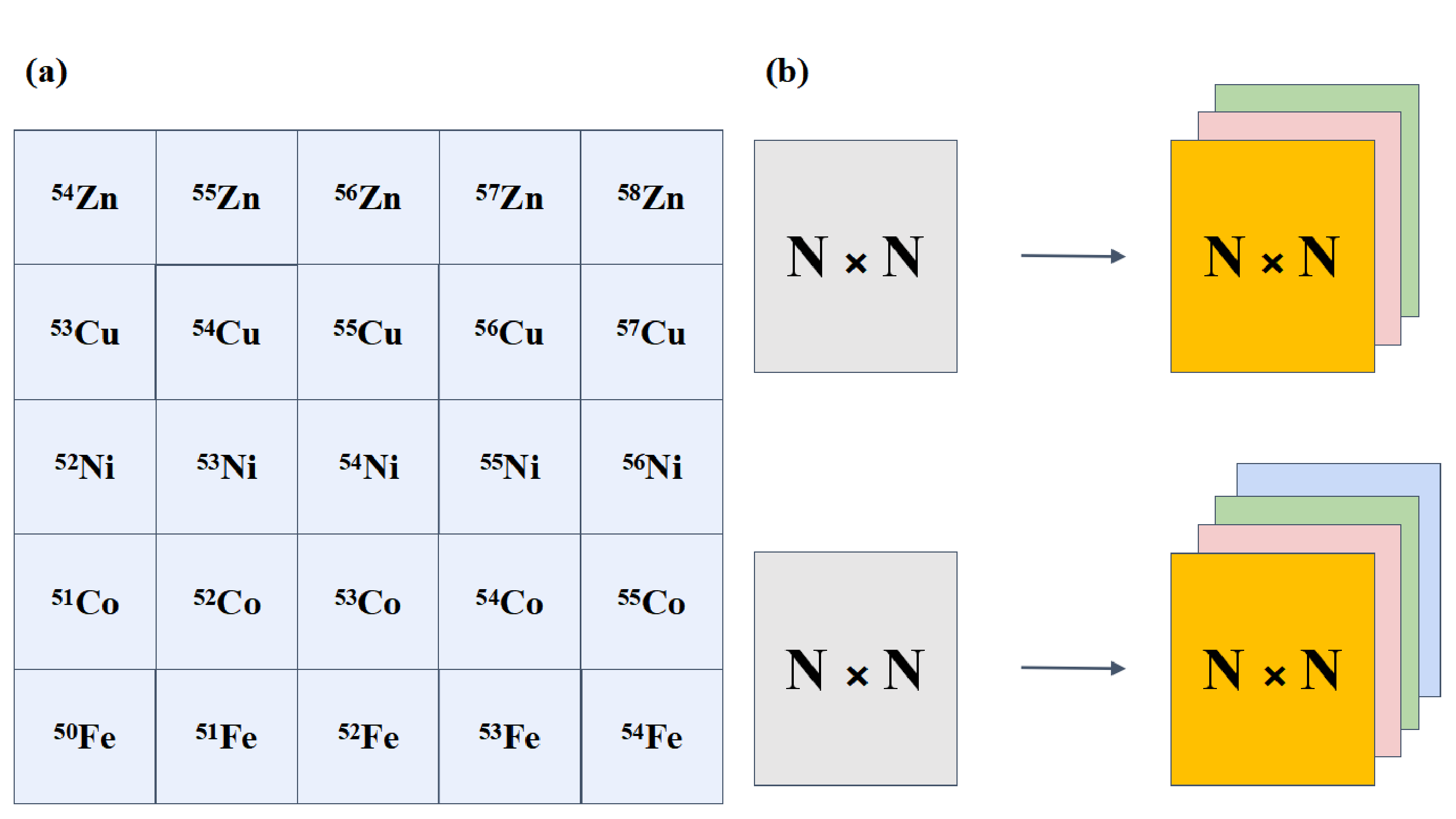}%
\caption{\label{fig:wide} (a) A $5\times5$ matrix containing information about the physics of 25 nuclei, where $^{54}\mathrm{Ni}$  is the target nucleus whose mass needs to be predicted, and the rest are its neighbors.
 (b) The upper subfigure displays three input channels (three different colors); the lower subfigure displays four input channels (four different colors).}\label{fig:preprosess}
\end{figure*}

Many nuclear mass models are nowadays based on global theories. The global theory can effectively capture the systematic changes in nuclear ground-state properties and excels in describing and predicting the physical quantities of nuclei in the regions where experimental data are limited. Conversely, the local theories based on experimental data emphasize the specific characteristics of ground-state properties within local contexts and have advantages in describing known experimental data. To improve predictive ability, we combine the global theoretical frameworks with the methods for extracting local relationships. The CNN is a kind of deep feedforward neural network with local connection and weight sharing. The data features around the target nucleus can be efficiently extracted by CNN, which is the reason why we use the CNN method. The convolutional layer is the main part of CNN. The function of the convolutional layer is to extract the features of a local region, which plays a key role in obtaining the local physical information of the target nucleus. Our objective is to predict the mass of a nucleus composed of $Z$ protons and $N$ neutrons. Drawing inspiration from the generalized Garvey-Kelson local mass relation~\cite{zhaoyvminlocal0}, we have devised the convolutional neural network inputs with a $5\times5$ dimension for each channel. In Fig.~\ref{fig:preprosess}(a), taking $^{54}\mathrm{Ni}$ as an example, we illustrate a $5\times5$ matrix containing information about the physics of 25 nuclei, where $^{54}\mathrm{Ni}$  is the target nucleus whose mass needs to be predicted, and the rest are its neighbors. 

In Fig.~\ref{fig:preprosess}(b), the upper subfigure illustrates that the inputs consist of three channels ($Z$, $N$, $M$): the first two channels represent the number of protons ($Z$) and the number of neutrons ($N$), while the third channel corresponds to the mass of the nucleus being predicted, designated as zero, with the masses of the neighboring nuclei ($M$)  serving as inputs (this model is referred to as CNN-I3
for short). There is no corresponding experimental data in some regions around the nuclei at the edge of the nuclide chart, and these missing experimental data are uniformly replaced by the average of the total nuclear masses. Building upon the three input channels, we additionally incorporate the nuclear pairing effect ($\delta$) as the fourth input channel (this model is referred to
as CNN-I4 hereafter), which is illustrated in the lower subfigure. $\delta$ is a physical quantity related to the nuclear pairing effect:
\begin{equation}
    \delta=[(-1)^{N}+(-1)^{Z}]/2.
\end{equation}

Next, we present the details of the convolutional neural network. The size of the convolution kernel is set as $3\times3$ with a stride of 1. Each convolutional layer contains 72 convolutional kernels. To enhance the extraction of edge features, zero padding is applied at both ends. The two-dimensional convolution formula is stated as
\begin{equation}
    O(u,v)=\sum_{i=1}^{3}\sum_{j=1}^{3}g(i,j)h(u-i,v-j),
\end{equation}
where $O (u,v)$ denotes the value of the element at coordinates $(u,v)$ in the output matrix and $g(i,j)$ denotes the value of the element at the coordinates $(i,j)$ in the convolution kernel. Similarly, $h(u-i,v-j)$ represents the value of the element at the coordinates $(u-i,v-j)$ in the input matrix.

There is no pooling layer in our network structure. It consists of two convolutional layers, each followed by a $\rm{ReLU}$ activation function. The $\rm{ReLU}$ function is defined as
\begin{equation}
    \rm{ReLU}(\it{x}) = \begin{cases}
    x\ \ \   x\geqslant0, \\
    0\ \ \   x<0. \\
    \end{cases} 
\end{equation}
Neurons utilizing the $\rm{ReLU}$ activation function only require addition, multiplication, and comparison operations, thereby enhancing computational efficiency. The $\rm{ReLU}$ function is a left-bounded activation function, with a derivative of 1 when $x > 0$. This characteristic partially mitigates the issue of gradient vanishing in neural networks and accelerates the convergence rate.

For regression problems, the root-mean-square error (RMSE) is commonly selected as the loss function, expressed as:
\begin{equation}
    \rm{Loss}=\sqrt{\frac{1}{\it{N}_{\it{t}}}\sum^{\it{N}_{\it{t}}}_{i=1}(y_{\rm{pre}}-y_{\rm{tar}})^{2}},
\end{equation}
where $N_{t}$ represents the size of the training set, and $y_{\rm{pre}}$ represents the predictions of the CNN, and $y_{\rm{tar}}$ represents the target values. The loss function measures the difference between the outputs of the neural network and the training data. Network training would modify the network parameters using suitable optimization algorithms (optimizers) to minimize the loss function until the specified criteria are satisfied, resulting in a data-trained neural network.

In the final layer of the network, a fully connected layer is appended to construct a convolutional neural network with only one node in the output layer, tailored for regression tasks. The output of this node represents the regression prediction result. The output layer is divided into two types: one directly represents the experimental nuclear mass as the output, while the other utilizes the difference between the experimental nuclear mass and the theoretical nuclear mass as the output. The difference $\Delta M$ between the experimental nuclear mass and the theoretical nuclear mass can be expressed as:
\begin{equation}
    \Delta M=M_{\rm{exp}}-M_{\rm{th}}.
\end{equation}
The theoretical masses $M_{\rm{th}}$ are taken from the mass table WS4~\cite{wang2014surface}, which is one of the most accurate nuclear mass tables. The experimental nuclear mass data are taken from AME2020 and AME2016~\cite{wang2021ame,wang2017ame2016}. To provide reliable predictions in the present work, we use all the existing data, and do not distinguish between the measured and extrapolated masses compiled in the mass evaluations (AME2016 and AME2020). We selected the masses of nuclei with $Z\geqslant 8$ and $N\geqslant 8$ that had already appeared in AME2016 as the training set. The newly appeared nuclear masses in AME2020 are used for extrapolation. 

\section{\label{sec:3}Results and discussion}

\begin{figure}
\centering
\includegraphics[width=8cm]{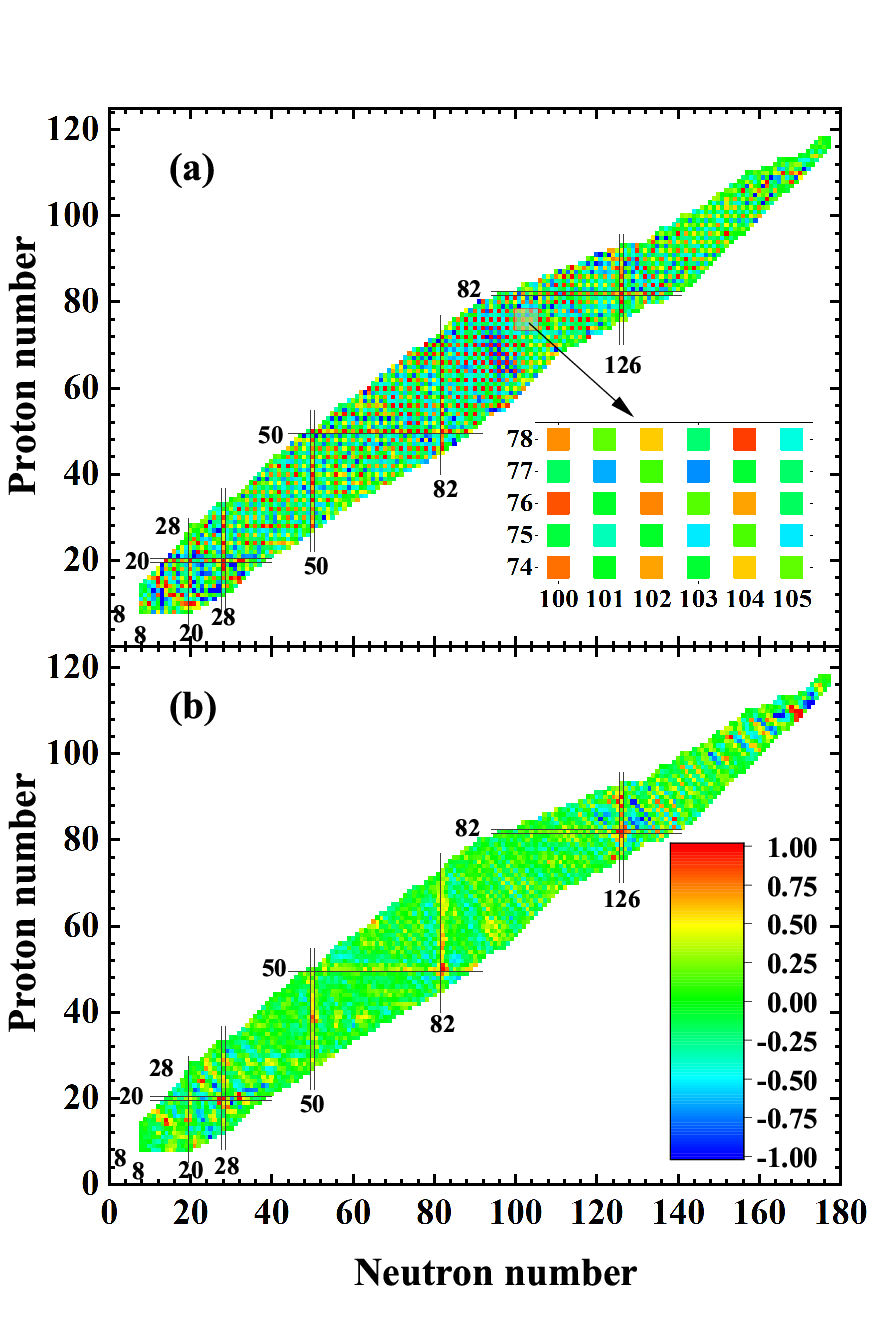}
\caption{\label{fig:wide}Differences (in MeV) between the nuclear masses predicted by CNN models and the experimental nuclear masses. (a) CNN-I3 model, RMSE: 0.535 MeV; (b) CNN-I4 model RMSE: 0.291 MeV.}\label{fig:direct}
\end{figure}

By designing the input data, the CNN-I3 and CNN-I4 models are constructed, with RMSE values of 0.535 MeV and 0.291 MeV on the training set, respectively. As the RMSE can only capture the general deviations between theoretical and experimental data, for a more detailed analysis of each nucleus, the differences between the predicted nuclear mass of the two CNN models and the experimental nuclear mass on the nuclide chart are presented in Fig.~\ref{fig:direct}. Figure~\ref{fig:direct}(a) presents the results of the CNN-I3 model. The results demonstrate a distinct odd-even staggering. The difference in each nucleus varies from that of its neighboring nuclei. This variation is closely related to the odd-even nature of the nucleus's proton and neutron numbers. From the subgraph, it can be further found that when the neutron number is even, the nuclear masses obtained by the CNN-I3 model are larger than the experimental nuclear masses. When the neutron number is odd, the nuclear masses obtained by the CNN-I3 model are smaller than the experimental nuclear masses. For the proton number, the same pattern holds true as for the neutron number. Therefore, Fig.~\ref{fig:direct}(a) exhibits a distinct pattern, inspiring us to incorporate additional physical conditions. Consequently, the pairing effect is further introduced in the input layer, resulting in the expansion of the three-input channel to a four-input channel, namely the CNN-I4 model. 

The CNN-I4 model achieves an overall RMSE of 0.291 MeV, representing a 46\% decrease compared to the CNN-I3 model. After introducing the pairing effect, the prominent odd-even staggering observed in the results of the CNN-I3 model almost disappears. Figure~\ref{fig:direct}(b) illustrates the differences between the masses predicted by the CNN-I4 model and the experimental nuclear masses. The overall prediction ability is significantly enhanced upon the inclusion of the pairing effect.

For the CNN-I4 model, the predicting accuracy is high even in the vicinity of magic numbers. The RMSE values of the CNN-I3 model for the Ca, Ni, Sn, and Pb isotopic chains are 0.798, 0.617, 0.812, and 0.652 MeV, respectively. In contrast to CNN-I3, the CNN-I4 model shows significantly reduced RMSE values in these four isotopic chains, with values of 0.487, 0.284, 0.436, and 0.409 MeV, respectively. It is evident that the model's predicting accuracy can be enhanced through the introduction of more physical information via a well-considered network input design. However, a significant deviation near the double magic number is also observed, particularly in the two isotopic chains of Sn and Pb. 

From Fig.~\ref{fig:direct}(b), it can also be observed that the predictive capability of the CNN-I4 model is poor in both the light and heavy nuclei regions. The model could be enhanced by incorporating additional physical information. In the above work, CNN is employed to extract local features from input data by using convolutional kernels to perform sliding operations. However, a series of nuclear mass physics models developed within the framework of global theory have not yet been considered. Next, our CNN model is comprehensively considered in combination with the global nuclear mass model.

\begin{figure}
\centering
\includegraphics[width=8cm]{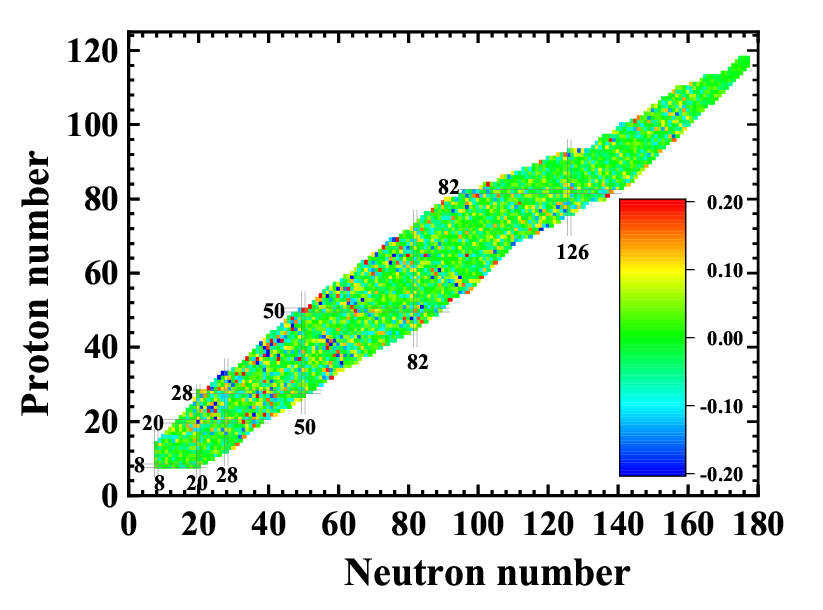}
\caption{\label{fig:wide}Differences (in MeV) between the predicted nuclear masses of CNN-WS4 model and the experimental nuclear masses.}\label{fig:CNN-WS4}
\end{figure}

Based on the CNN-I4 model, a specialized design of the output layer is employed to predict the differences between the WS4 nuclear mass~\cite{wang2014surface} and the experimental nuclear mass. This leads to the development of a new model that integrates global and local theoretical frameworks (this model is referred to as
CNN-WS4 for short). The differences between the predicted nuclear masses of the CNN-WS4 model and the experimental nuclear masses are illustrated in Fig.~\ref{fig:CNN-WS4}. Note that the scale of color used in Fig.~\ref{fig:CNN-WS4} is smaller than that used in Fig.~\ref{fig:direct} by a factor of 5. The WS4 nuclear mass model is incorporated into the convolutional neural network, resulting in a significant improvement in overall predicting ability. The RMSE within the known experimental data range decreased to 0.070 MeV. This represents the highest accuracy achieved so far in theoretical models for predicting nuclear masses, compared with the existing studies~\cite{jiang2012predictions,WUxinghuikrr128kev,niu2022nuclear}. The result proves the validity of the CNN-WS4 approach, which combines the global theoretical
model with methods capable of extracting local features to predict nuclear mass. 

An interesting phenomenon is that in the region of superheavy nuclei ($Z\geqslant104$), the CNN-WS4 model predicts a deviation of 0.046 MeV from the experimental masses. This indicates that the CNN-WS4 model also shows superior predictive capability in the region of superheavy nuclei. As a result, a global-local model (CNN-WS4) has been developed through the careful design of the network's inputs and outputs, leading to higher predictive accuracy.

To evaluate the extrapolation performance of CNN-WS4, we make extrapolation using newly emerged nuclei from AME2020. The RMSE of the extrapolation realm is 0.211 MeV. Since many newly emerged nuclei in AME2020 are located at the edges of the nuclide chart, the constructed convolutional neural network model is unable to effectively extract the surrounding features of these nuclei, resulting in an overall increase in extrapolation deviation. 

\begin{figure*}
\centering
\includegraphics[width=16cm]{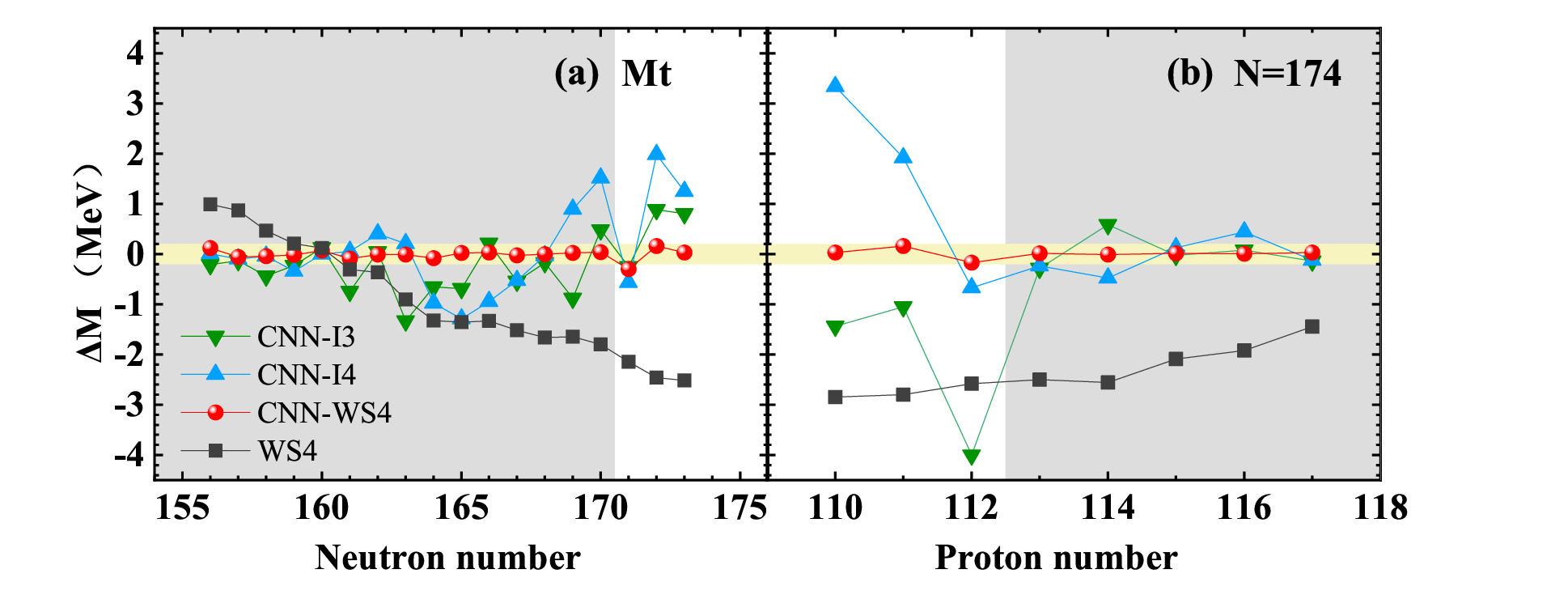}
\caption{\label{fig:wide} The learning and extrapolation mass differences $\Delta M$ (in MeV) of the CNN-I3 (green downward triangles), CNN-I4 (blue upward triangles), and CNN-WS4 models (red dots), with the WS4 nuclear mass model (black squares) as a reference. The shade (white) regions indicate the learning (extrapolation) areas. (a) The Mt isotopic chain; (b) $N=174$ isotonic chain.}\label{fig:extrapolation}
\end{figure*}

To demonstrate the capabilities of the CNN-WS4 model explicitly, we present in Fig.~\ref{fig:extrapolation} the learning and extrapolation performance of various models for the Mt isotopic chain and the $N=174$ isotonic chain. Using WS4 model to predict nuclear mass leads to significant deviations. On the Mt isotopic chain, the difference between the nuclear mass predicted by the WS4 model and the experimental nuclear mass varies from 0.994 MeV to $-2.519$ MeV as the neutron number increases. On the $N=174$ isotonic chain, the WS4 model exhibits a systematic deviation, with a difference between its predicted nuclear mass and the experimental nuclear mass ranging from $-2.845$ MeV to $-1.444$ MeV. Similarly, the predictive performance of only using convolutional neural networks is not satisfactory. 
In the learning region, the results of the CNN-I3 model and CNN-I4 model are significantly better than those of the WS4 model. However, when it comes to extrapolation, using CNN to predict nuclear mass shows noticeable fluctuations. The integrated CNN-WS4 model, combining these two approaches, demonstrates the best results in both the learning and extrapolation domains. Along these two chains, the CNN-WS4 model yields mass deviations within 0.2 MeV for the majority of nuclei, as highlighted by the yellow region. It can be concluded that considering more physical factors, especially the consideration of nuclear mass models, has led to significant improvements in both learning and extrapolation performance. In addition, it should be noted that the masses of most nuclei present in Fig.~\ref{fig:extrapolation} have not been measured, and further measurements of more nuclei are expected in the future to validate the model's ability.

\section{\label{sec:4}SUMMARY AND PERSPECTIVES}

In this study, we combine the global theoretical model with methods for extracting local relationships, aiming to enhance the predictive accuracy of nuclear mass. A convolutional neural network has been employed to successfully predict nuclear mass. To begin with, three kinds of physical information are considered in the input layer of the network: the proton number, the neutron number, and the mass of neighboring nuclei. 
To eliminate the observed odd-even staggering, the pairing effect of the nucleus is further incorporated into the input layer. Thoughtful design enhances the predictive power of our model significantly. Building upon this foundation, to maximize the utility of the theoretical model, we have specially designed the output layer and introduced the WS4 model. 

By considering a global nuclear mass model and utilizing CNN to extract features from surrounding nuclei, we have achieved a nuclear mass prediction accuracy of 0.070 MeV. The CNN-WS4 model also shows excellent predictive
capability in the realm of superheavy nuclei. When conducting extrapolation, this model also exhibits appreciable stability, thereby demonstrating its robustness. This indicates that CNN has an advantage in predicting nuclear mass. To further validate the stability of the CNN-WS4 model, we divide the data into a training set and a testing set in an 8:2 ratio. In AME2020, there are 3456 nuclei with $Z\geqslant 8$ and $N\geqslant 8$. A random selection of 691 nuclei is used as the testing set, while the remaining 2765 nuclei are utilized as the training set. The RMSE for the training set is 0.095 MeV, while the RMSE for the testing set is 0.171 MeV. This result further demonstrates the robustness of the CNN-WS4 model.  

Currently, our research solely relies on the WS4 theoretical model. If more physical models are introduced into the design process of the network, what kinds of effects on the stability of the network and the reliability of the prediction results still need to be further explored. It is also an open question about the interpretability of neural networks. Physical information has a great influence on the stability of neural networks and the reliability of prediction results. In the future, the development of neural network methods that consider a greater number of physical effects or constraints is expected to further enhance the predictive capability for nuclear mass. 

\vspace{\baselineskip}

\begin{acknowledgments}
This work was supported by the Natural Science Foundation of Jilin Province (No. 20220101017JC), National Natural Science Foundation of China (No. 11675063, No. 12375109, No. 11875070, and No. 11935001), Key Laboratory of Nuclear Data Foundation (JCKY2020201C157), the Anhui project (Z010118169), the Key Research Foundation of Education Ministry of Anhui Province under Grant No. 2023AH050095, the JSPS Grant-in-Aid for Scientific Research (S) under Grant No. 20H05648, the RIKEN iTHEMS Program, and the RIKEN Pioneering Project: Evolution of Matter in the Universe.
\end{acknowledgments}

\nocite{*}

\bibliography{apssamp}

\end{document}